# Contribution of subradiant plasmon resonance to electromagnetic enhancement in resonant Raman with fluorescence examined by single silver nanoparticle dimers

Tamitake Itoh[1]*, Yuko S. Yamamoto[2]

[1]Nano-Bioanalysis Research Group, Health Research Institute, National Institute of Advanced Industrial Science and Technology (AIST), Takamatsu, Kagawa 761-0395, Japan

[2]School of Materials Science, Japan Advanced Institute of Science and Technology (JAIST), Nomi, Ishikawa 923-1292, Japan

*Corresponding author: tamitake-itou@aist.go.jp

## ABSTRACT

Here, we investigate the spectral relationships between electromagnetic (EM) enhancement and surface-enhanced resonant Raman scattering (SERRS) with surface-enhanced fluorescence (SEF), which is background emission of SERRS, in the context of light-matter interactions, using single silver nanoparticle dimers as a model system. We focus on the lowest energy (superradiant) plasmon in far-field scattering to examine EM enhancement. We classify the spectral relationships into two types: those in which the spectral envelopes of SERRS with SEF have spectral shapes similar to those of plasmon resonance and those in which the spectral envelopes of SERRS with SEF exhibit higher energy shifts than the plasmon resonance. By examining these results, we aim to determine the degree of morphological asymmetry in the dimers based on an EM mechanism. Our analysis of the two types of spectral relationships reveals that dipole–dipole and dipole–quadrupole coupled plasmon resonance (subradiant Fano resonance) are responsible for EM enhancement.

## I. INTRODUCTION

The Purcell effect largely enhances the effective interaction cross sections between light and molecules near plasmonic metal nanostructures [1]. In particular, the Raman scattering of a molecule located inside a nanogap or junction of a plasmonic nanoparticle (NP) dimer exhibits an enhancement factor of up to $10^{10}$, which enables single-molecule Raman spectroscopy under the resonant Raman condition [2-5]. This phenomenon is called surface-enhanced resonant Raman scattering (SERRS), and the location where it occurs is called a hotspot (HS). In the case of a dye molecule, fluorescence exhibits both enhancement and quenching at the HS. This phenomenon is called surface-enhanced fluorescence (SEF), which appears as background emission in SERRS [1]. The HSs of SERRS and SEF have received considerable attention because they exhibit exotic phenomena beyond conventional classical spectroscopies, such as strong coupling, ultrafast ultra-fast SEF, vibrational pumping, and the field gradient effect [1,6–13]. The origin of these phenomena is an extremely small plasmonic mode volume of approximately several cubic nanometers, which leads to a large amplitude in both incident light and vacuum fluctuation [1,7,8]. Therefore, it is important to quantitatively clarify the relationship between these phenomena and plasmon resonance to control or optimize them.

The relationship between SERRS, SEF, and plasmon resonance has been studied to quantitatively clarify these phenomena based on an electromagnetic (EM) mechanism [14,15]. The EM enhancement of SERRS with SEF is described as the product of an excitation enhancement factor $F_{\mathrm{R}}(\omega_{\mathrm{ex}})$ and an emission enhancement factor $F_{\mathrm{R}}(\omega_{\mathrm{em}})$ caused by plasmon resonance and can be expressed as follows:

$$F_{\mathrm{R}}(\omega_{\mathrm{ex}},\mathbf{r})F_{\mathrm{R}}(\omega_{\mathrm{em}},\mathbf{r}) = \left|\frac{E_{\mathrm{loc}}(\omega_{\mathrm{ex}},\mathbf{r})}{E_{\mathrm{i}}(\omega_{\mathrm{ex}})}\right|^{2} \times \left|\frac{E_{\mathrm{loc}}(\omega_{\mathrm{em}},\mathbf{r})}{E_{\mathrm{i}}(\omega_{\mathrm{em}})}\right|^{2}, \qquad (1)$$

where $E_I$ and $E_{loc}$ indicate the amplitudes of the incident and enhanced local electric fields, respectively; $\omega_{ex}$ and $\omega_{em}$ denote the frequencies of the incident and Raman or fluorescent light, respectively; and **r** is the position of a molecule in an HS [4]. We have conducted experiments to demonstrate that the spectral envelopes of SERRS with SEF have shapes that resemble those of plasmon resonance spectra observed in far-field elastic scattering [14,15]. This correlation between spectra is consistent with the EM mechanism, which predicts the spectral modulation in SERRS and SEF by the factor $F_R(\omega_{em})$ in Eq. (1) [15]. However, there has been disagreement over the spectral correlation between SERRS and SEF and plasmon resonance spectra. Some studies have reported SERRS spectra that do not align with plasmon resonance spectra, as observed in far-field elastic scattering or extinction spectra [16,17]. These findings suggest that the spectral shapes of plasmon resonance deviate from the $F_R(\omega_{em})$ in Eq. (1). The contribution of the subradiant plasmon resonance, which is not visible in far-field scattering spectra, to $F_R(\omega_{em})$ has been debated [17-19]. Therefore, this controversy has to be resolved to accurately understand the mechanism of SERRS and SEF in HSs.

The roles of subradiant plasmon resonance in photo-induced phenomena, such as plasmon-induced hot electrons, magnetic resonance enhanced Raman scattering, and photoluminescence, have been reported in several studies [20,21,22]. These studies have also highlighted the significance of the large $F_R$ of subradiant plasmon resonance due to the suppression of radiative decay loss. The mechanism of the enhancement in photo-induced phenomena has been described by the coupling energy between radiant and subradiant plasmon modes [23,24].

In this study, we examined the spectral relationships between SERRS, SEF, and plasmon resonance in single silver NP dimers. We focused on the lowest energy plasmon resonance, which corresponds to the superradiant dipole resonance. The spectral envelopes of the SERRS with SEF were mostly similar to those of the plasmon resonance spectra. However, we also observed some dimers with envelopes that shifted to higher-energy regions from the plasmon resonance maxima. There was significant dimer-to-dimer variability in the degree of these shifts. The SERRS with SEF spectra with the highest energy shifts had envelope maxima around the dips in plasmon resonance spectra. To better understand these results, we compared the SERRS and SEF spectra to SEM images of the dimers. This revealed that the morphological asymmetry is related to higher energy shifts. We used numerical calculations based on the EM mechanism to analyze the effect of dimer morphology on the spectral relationships. Our analysis of the calculated electric near and far fields and their phases at the HSs showed that the first type of spectral envelope, which is similar to the plasmon resonance spectra, is mainly caused by dipole–dipole coupled plasmon resonance in symmetric dimers. The second type, in which the spectral envelopes shift to higher energies, is mainly caused by dipole–quadrupole coupled plasmon resonance (Fano resonance) in asymmetric dimers, resulting in deviation from the plasmon resonance spectra. These results will be useful in various plasmon-enhanced spectroscopies [25].

## II. EXPERIMENT

Colloidal silver NPs with a mean diameter of approximately 35 nm ($1.10\times10^{-10}$ M) were prepared for the SERRS experiment using the Lee and Meisel method [26]. The colloidal silver NP dispersion was mixed with an equal volume of R6G aqueous solutions ($1.28\times10^{-8}$ M) containing 5 mM NaCl and allowed to aggregate for 30 minutes to generate SERRS activity. The final concentrations of the R6G solutions ($6.34\times10^{-9}$ M) and NP dispersion ($5.5\times10^{-11}$ M) were within a single molecular SERRS condition, as confirmed by a two-analyte or isotope technique [27,28].

A sample solution of 50 μL was then dropped onto a glass slide plate and sandwiched between a glass cover plate to immobilize the SERRS-active colloidal silver NPs. The sample was left on the plate to allow the NPs to stably attach to the glass slide plate.

The sample plate was placed under an inverted optical microscope (IX-71; Olympus, Tokyo, Japan).

Figures 1(a1) and 1(b1) show the elastic light scattering and SERRS with SEF measurements of the same sample area on a glass plate surface. Figures 1(a2) and 1(b2) show images of elastic light scattering and SERRS with SEF, respectively. The elastic scattering light from single silver NP dimers was detected by illuminating unpolarized white light from a 50-W halogen lamp through a dark-field condenser (numerical aperture (NA) 0.92). The NA of the objective lens (LCPlanFL 100×, Olympus, Tokyo) was set to 0.6 for dark-field illumination when measuring elastic scattering light. SERRS with SEF light from single silver NP dimers was measured by illuminating an unpolarized polarized excitation green laser beam (2.33 eV (532 nm), 3.5 W/cm$^2$, Depolarizer DEQ-2S SigumaKoki) from a CW Nd3+: YAG laser (DPSS 532, Coherent, Tokyo) on the sample plate through another objective lens (5×, NA 0.15, Olympus, Tokyo). The NA of the objective lens was increased to 1.3 when measuring the SERRS with SEF light to efficiently collect the emitted light. Single NP elastic scattering and SERRS with SEF spectra were measured by selecting a spot on the image plane using a pinhole in front of a polychromator equipped with a thermoelectrically cooled charge-coupled device (CCD) assembly (DV 437-OE-MCI, Andor, Japan). The SERRS-active silver NPs were aggregated [14], and if the selected SERRS-active silver NP aggregates showed dipolar plasmon resonance with maxima of 1.7–2.1 eV, they were dimers [14]. Colloidal gold NPs (mean diameters of 60, 80, and 100 nm; EMGC40, Funakoshi, Japan) were used to convert the scattering intensities into cross-sections. The detailed procedure for this conversion is provided elsewhere [29].

In this study, we treated SERRS and SEF spectra in the same manner because both are equally modulated by $F_R(\omega_{em})$ in Eq. (1) [1]. A strict separation of both spectra under the framework of the EM mechanism is provided in the Supporting Information (Sec. 1). In short, the intensity instability of SEF is more sensitive to molecular fluctuations inside the HS than that of SERRS, due to the quenching factor for SEF.

## III. RESULTS AND DISCUSSION

We examined the relationship between the plasmon resonance, SERRS and SEF in single dimers. Figs. 2(a)–2(c) show four different variations in the spectral relationships. The plasmon resonance appears as a maximum in the elastic scattering cross-section ($\sigma_{sca}(\omega)$) spectra at energies in the range 1.6–2.0 eV of $\hbar\omega$, where $\omega$ is the angular frequency of light. Figs. 2(a1)–2(a4) show that the maxima of the SERRS with SEF spectra (red lines) are similar to the plasmon resonance maxima (blue lines). This spectral correlation is a dominant feature in the current SERRS-active dimers and is consistent with the EM mechanism [1,14,15]. In the EM mechanism of SERRS and SEF, the Raman and fluorescence processes are enhanced by a factor of two due to the plasmon resonance, as indicated by $F_R(\omega_{ex})$ and $F_R(\omega_{em})$ in Eq. (1). This leads to a modulation of the spectral envelopes of the SERRS with SEF spectral envelopes by $F_R(\omega_{em})$ [14,15]. More information on the spectral modulation by $F_R(\omega_{em})$ via EM enhancement can be found in the Supporting Information (Section 1). However, the SERRS with SEF spectra do not show this spectral correlation. Figures 2(b1)–2(b4) show that the SERRS with SEF spectra exhibit higher energy shifts from the plasmon resonance maxima, with considerable variation between dimers. Figures 2(c1)–2(c4) show that the SERRS with SEF spectra have maxima around the dips in the $\sigma_{sca}(\omega)$ spectra at around 2.2 eV.

We examined the relationship between the plasmon resonance maxima in the $\sigma_{sca}(\omega)$ spectra and the spectral envelopes of the SERRS and SEF for 39 dimers. As shown in Fig. 3(a), two trends were observed in this relationship. The first trend, indicated by the black open circles, is a proportional correlation between $\hbar\omega_P$ (plasmon resonance maximum energy) and $\hbar\omega_s$ (spectral maximum energy). We refer to this correlation as "Type I." The second trend, indicated by the red open circles, is that the values of $\hbar\omega_s$ remain around 2.1 eV; even as the values of $\hbar\omega_P$ vary widely from 1.85 to 2.2 eV. We refer to this lack of correlation as "Type II." There is also an intermediate type between Types I and II. The fact that two types are commonly observed in the relationships between $\hbar\omega_P$ and $\hbar\omega_s$ suggests that there are two types of dimers emitting SERRS with SEF light. We propose that the dimers responsible for Type I generate enhanced EM field directly through superradiant plasmon resonance. Conversely, the dimers that cause Type II may generate EM enhancement through subradiant plasmon resonance and emits SERRS with SEF light to free space through superradiant resonance that is coupled with subradiant resonance [30,31]. One possible candidate for subradiant plasmons is higher-order plasmons, such as quadrupole and hexapole plasmons, which can appear in the $\sigma_{sca}(\omega)$ spectra of large dimers in the visible region due to the retardation effect [32]. These dimers exhibit large values of $\sigma_{sca}(\omega)$ because $\sigma_{sca}(\omega)$ is approximately proportional to the square of the dimer volume [32].

To further explore the differences between Type I and Type II, we compared the $\sigma_{sca}(\omega)$ at $\hbar\omega_P$ for each type. As shown in Figure 3(b), there were no clear differences in the distributions of Types I and II, suggesting that dimer size is not a significant factor in determining the two different spectral trends. This result indicates that dimer shape, rather than size, may be an important factor in the formation of Types I and II.

To understand the origin of Types I and II, we investigated the relationship between dimer shapes, SERRS with SEF spectra, and plasmon resonance spectra. Figs. 4(a) and 4(b) show SEM (JSM-6700F, JEOL) images of dimers on ITO glass, along with their SERRS with SEF spectra (red lines), and plasmon resonance spectra (blue lines). The details of measurement procedures are described elsewhere [14]. The smaller values of $\sigma_{sca}(\omega)$s in Figs. 4(a) and 4(b) compared to those in Fig. 2 are due to the fact that the dimers are placed at the interface between air and ITO glass plate, whereas the dimers in Fig. 2 were placed at the interface between water and a glass plate. Figs. 4(a1)–4(a4) and 4(b1)–4(b4) illustrate the spectral correlation and non-correlation between SERRS with SEF and plasmon resonance, respectively. The ratios of the two NP diameters, $D_1/D_2$ (as shown in Fig. 5(b)), are approximately 1.0 to 1.3 and approximately 1.8 to 2.5 for Figs. 4(a1)–4(a4) and Figs. 4(b1)–4(b4), respectively. It indicates that Figs. 4(a1)–4(a4) and Figs. 4(b1)–4(b4) show morphologically symmetric and asymmetric dimers, respectively. Therefore, these results indicate that the morphological symmetry and asymmetry of the dimers resulted in spectral correlation and non-correlation, respectively. We plan to evaluate the effect of junction morphology using SEM with higher spatial resolution.

To further understand the properties of the spectral relationships between superradiant plasmon resonance and SERRS with SEF, as presented in Figs. 2–4, we analyzed the experimental results using electromagnetism and the finite-difference time-domain (FDTD) method (EEM-FDM Ver.5.1, EEM Co., Ltd., Japan). The key findings are summarized as follows:

(1) There are two spectral relationships (Types I and II) between superradiant resonance and SERRS with SEF. Type I exhibits an envelope maximum of SERRS with a SEF at a position similar to the superradiant resonance maxima. Type II exhibits an envelope maxima of ~2.1 eV around the dips in the $\sigma_{sca}(\omega)$ spectra.

(2) The intermediate type between Types I and II is continuous, with the envelope maxima of SERRS with SEF existing continuously between the superradiant resonance maxima and dips in the $\sigma_{sca}(\omega)$ spectra.

(3) There is no clear relationship between the values of $\sigma_{sca}(\omega_p)$ at superradiant resonance maxima and Types I or II, indicating that dimer size is not a significant factor in distinguishing Type I and II.

(4) Symmetric and asymmetric dimers are likely associated with Types I and II, respectively.

To confirm these four properties, we used FDTD calculations. The complex refractive index of silver for NP dimers was taken from Ref. 33. The effective refractive index of the surrounding medium was set at 1.39 from the maximum energy of a gold NP with a diameter of 80 nm [29]. The validation of the calculation conditions for reproducing the experimental conditions has been described elsewhere [14,29]. Figures 5(a) and 5(b) illustrate the coordinate system of a dimer and the setup for exciting a dimer composed of two spherical NPs with diameters $D_1$ and $D_2$, respectively, while maintaining a gap of 1 nm. We calculated the $\sigma_{sca}(\omega)$ spectra (shown in Fig. 5(c)) and the $F_R(\omega)$ spectra (shown in Fig. 5(d)) using the FDTD method. The $\sigma_{sca}(\omega)$ spectra represent the scattering cross-section of the dimers, whereas the $F_R(\omega)$ spectra represent the field enhancement factor at the gap between the two NPs. In our calculations, we used excitation polarization parallel to the dimer long axis, with a phase of local electric field ($E_{loc}$) at the center gap set to 0 and phase of incident electric ($E_I$) set to 180°. We reproduced the $\sigma_{sca}(\omega)$ and $F_R(\omega)$ spectra for unpolarized light excitation by averaging the parallel and perpendicular excitation spectra. In the case of a NP dimer on a glass substrate, the parallel polarization component of the illumination light against the substrate surface can generate enhanced electric fields at the hotspot. As a result, the scattering spectra of dimer HSs obtained using conventional dark-field illumination with a low-NA condenser lens are insensitive to the illumination angles. Therefore, we assumed that the illumination polarization was parallel to the substrate surface in our calculations. This approximation may not be applicable for the HSs between plasmonic NPs on a plasmonic substrate, as the scattering spectra can be sensitive to the illumination angles in this case [34]. We defined the maximum energy in $F_R(\omega)$ spectrum as $\hbar\omega_F$, which corresponds to the experimentally obtained $\hbar\omega_S$, as indicated in Fig. 5(d). In our calculations, we did not consider the non-local effect [35], which can significantly decrease both SERRS and SEF intensity by creating electron-hole pairs. This effect does not significantly affect the spectral shape.

We previously discussed the origin of Types I and II shown in Fig. 3 using superradiant and subradiant plasmon resonance, respectively. The superradiant resonance generating SERRS and SEF at a HS is the dipole resonance; more precisely, it is the lower branch of the dipole–dipole (DD) coupled plasmon resonance of the two NPs [14,15]. Here, we explore the coupled plasmon resonance arising from subradiant resonance. Figures 6(a1)–6(a10) show the diameter dependence of the $\sigma_{sca}(\omega)$ spectra for spherical silver NPs calculated using Mie theory [32]. As the NP diameter increases, the dipole resonance in the $\sigma_{sca}(\omega)$ spectra shifts to the lower energy region and quadrupole resonance begins to appear in the higher energy region, as shown in Fig. 6(a6). It then shifts to the lower energy region, as shown in Fig. 6(a9). Figures 6(b1) and 6(b2) illustrate the surface charge distributions of a dipole without a retardation effect, and a quadrupole with a retardation effect, respectively. The presence of dipoles and quadrupoles suggests that two types of dimer geometry candidates could cause subradiant resonance. One candidate is the dipole–quadrupole (DQ) coupled resonance of a dimer with symmetric geometry; more precisely, the lower branch of the DQ coupled plasmon resonance [18]. Figure 6(c1) illustrates the surface charge distribution of a DQ-coupled plasmon of a symmetric dimer obtained by superposing dipole and quadrupole plasmons. This DQ-coupled resonance requires spectral overlap between the dipole and quadrupole resonances of each NP, requiring broad resonance linewidths of the dipole plasmon, as shown in Figs. 6(a7) to 6(a9). Another candidate for subradiant resonance is the DQ-coupled resonance of a dimer with asymmetric geometry. Figure 6(c2) illustrates the surface charge distribution of a DQ-coupled plasmon of an asymmetric dimer by superposing the dipole plasmon of the smaller NP and quadrupole plasmon of the larger NP [37]. Such DQ coupled resonance requires a spectral overlap between the dipole resonance of the smaller NP and quadrupole resonance of the larger NP. Therefore, the combination of NPs in Figs. 6(a2) and 6(a8) can generate a DQ-coupled resonance when the two resonances overlap.

We examined the possibility of DQ-coupled plasmon resonance of symmetric dimers (shown in Fig. 6(c1)) as a subradiant plasmon resonance causing Type II. Figures 7(a1)– 7(a6) show the NP diameter dependence of both the $\sigma_{sca}(\omega)$ and $F_R(\omega)$ spectra for the symmetric dimers with unpolarized light excitation. The values of $F_R(\omega)$ were obtained from HS, as shown in Fig. 5(b). As the diameter of NPs increases, both $\hbar\omega_P$ and $\hbar\omega_F$ spectra experience a shift toward lower energies. At the same time, the $F_R(\omega)$ spectra broadens to the higher energy side of $\hbar\omega_F$ ~2.0 eV, as shown in Fig. 7(a3). Finally, for an NP diameter of 120 mm, $\hbar\omega_F$ moves from the position of $\hbar\omega_P$ to a position around the dip in the $\sigma_{sca}(\omega)$ spectrum in Fig. 7(a6). The dip is the result of the destructive interference between DD and DQ coupled plasmons, indicating that the dip position corresponds to DQ coupled resonance [18]. The difference between $\hbar\omega_P$ and $\hbar\omega_F$ is also observed experimentally as the higher energy shifts in SERRS with SEF spectral envelopes, as shown in Fig. 3(a). These higher energy shifts in $\hbar\omega_F$ are observed for NP diameters of 120 nm, indicating that the values of $\sigma_{sca}(\omega_P)$ and $\hbar\omega_P$ of DD coupled resonance for Type II dimers should be $>>150 \times 10^{-3}$ $\mu m^2$ and $< $ ~1.6 eV, respectively, as shown in Fig. 7(a6). These values of $\sigma_{sca}(\omega_P)$ and $\hbar\omega_P$ are much larger and smaller, respectively, than their experimentally observed values, as shown in Fig. 3(b). In fact, Fig. 3(b) shows that a dimer with $\sigma_{sca}(\omega_P)$ and $\hbar\omega_P$ of ~5 $\times$ $10^{-3}$ $\mu m^2$ and 2.08 eV exhibits higher energy shifts. This large difference between the experimental and calculated values suggests that the symmetric dimer may not be a candidate for Type II.

Next, the DQ-coupled plasmon resonance of the asymmetric dimers, as shown in Fig. 6(c2), was evaluated as a potential candidate for Type II. Figures 7(b1)–7(b6) show the dependence of both the $\sigma_{sca}(\omega)$ and $F_R(\omega)$ spectra on $D_1$ at HS, with $D_2$ fixed at 30 nm, while using unpolarized light excitation. As $D_1$ increases, both $\hbar\omega_p$ and $\hbar\omega_F$ experience a shift to lower energies until $D_1$ reaches 80 nm, as shown in Figs. 7(b1)–7(b3). For $D_1 >$ 80 nm, $\hbar\omega_p$ continues to shift toward lower energies, but $\hbar\omega_F$ remains at approximately 2.2 eV, as shown in Figs. 7(b3)–7(b5). Finally, $\hbar\omega_F$ overlaps with the dip position in the $\sigma_{sca}(\omega)$ spectrum, which may be the DQ-coupled resonance position, as shown in Fig. 7(b6), for $D_1$ of 120 nm. This difference in trends between $\hbar\omega_p$ and $\hbar\omega_F$ is consistent with the experimentally observed higher-energy shifts in the spectra of SERRS with SEF, as shown in Figs. 2(b) and 2(c). Furthermore, the values of the calculated $\sigma_{sca}(\omega)$ and $\hbar\omega_p$ are reasonably consistent with the experimental values for Type II, as shown in Fig. 3(b). Therefore, we conclude that the asymmetry in the dimers causes a DQ-coupled resonance, resulting in Type II. In summary, the subradiant resonance discussed in the experimental section is the DQ-coupled resonance of asymmetric dimers. This discussion is consistent with Fig. 4, which shows the relationship between dimer shape, plasmon resonance spectra, and SERRS with the SEF spectra.

Figure 7 indicates that asymmetric dimers are responsible for spectral deviation in the DD-coupled plasmon resonance from the envelopes of SERRS with SEF based on the spectral relationships between DQ (or DD) coupled resonance and $F_R(\omega)$ at the HSs. To confirm this, we provide evidence of DQ (or DD) coupled resonance observed in the $\sigma_{sca}(\omega)$ and $F_R(\omega)$ spectra. The amount of phase retardation of $E_{loc}$ against the phase of $E_I$ is shown in Eq. (1). The $E_{loc}$ of the DD-coupled resonance induces a phase retardation of 90° as compared to $E_I$ [1]. The phase of $E_{loc}$ of the DQ-coupled resonance forces a further retardation of 90° from the DD-coupled resonance because the DQ-coupled resonance receives light energy by coupling with the DD-coupled resonance [30,31]. Thus, the $E_{loc}$ of the DQ-coupled resonance exhibits a phase retardation of 180° from $E_I$. The destructive interference between the $E_{loc}$ of DD-coupled resonance and the $E_{loc}$ of DQ-coupled resonance appears as a dip structure in the $\sigma_{sca}(\omega)$ spectrum.

We determined the relationship between the DD (or DQ) resonance and the $E_{loc}$ phase with excitation polarization parallel to the dimer long axis. Figures 8(a1)–8(a4) illustrate the $D_1$ dependence of $\sigma_{sca}(\omega)$ spectra on the phases of $E_{loc}$. The initial phase of $E_I$ was set to 180° and $D_2$ was set to 30 nm. The $\sigma_{sca}(\omega)$ spectra exhibit a phase retardation by 90° due to $E_{loc}$ at $\hbar\omega_p$, indicating that these spectral maxima correspond to the DD coupled resonance. As the value of $D_1$ increases, the DD-coupled resonance spectra become broader due to the increasing effect of radiation damping [38]. When $D_1$ is further increased, a dip structure appears in the $\sigma_{sca}(\omega)$ spectrum at a phase retardation of 180°, indicating that the spectral dip is caused by destructive interference between the DD and DQ coupled resonance.

Figs. 8(b1)–8(b4) show the $D_1$ dependence of the $F_R(\omega)$ spectra at an HS with the phase of $E_{loc}$. The $F_R(\omega)$ spectrum in Fig. 8(b1) exhibits $\hbar\omega_F$ at a phase retardation of 90°, indicating that the maximum is the result of the EM enhancement by the DD-coupled resonance. As $D_1$ increases, $\hbar\omega_F$ exhibits a lower energy shift, as shown in Figs. 8(b2). However, the lower energy shift in $\hbar\omega_F$ does not occur in Figs. 8(b3) and $\hbar\omega_F$ remains at approximately ~2.18 eV, as shown in Fig. 8(b4), where the phase retardation is 180°. This position is the same as the dip in the $\sigma_{sca}(\omega)$ spectrum shown in Fig. 9(a4), indicating that the $F_R(\omega)$ maximum at $\hbar\omega_F$ is the result of EM enhancement by the DQ-coupled resonance, which obtains light energy through DD-coupled resonance. This mechanism of EM enhancement by the DQ coupled resonance explains the experimentally observed higher-energy shifts in the envelope maxima of SERRS with SEF, as shown in Figs 2(b) and 2(c). This mechanism also explains why the envelope maximum positions of SERRS and SEF match the dip positions in the $\sigma_{sca}(\omega)$ spectra shown in Figs. 2(b) and 2(c).

In short, the experimentally observed spectral deviation of the envelope of SERRS with SEF from DD-coupled resonance is explained as the contribution of DQ-coupled resonance, which is subradiant to the EM enhancement. Fig. 9 summarizes the spectral relationship between $\sigma_{sca}(\omega)$, $F_R(\omega)$, and the phase of $E_{loc}$ at the HSs for the asymmetric dimers and clarifies the mechanism of the spectral deviation.

Figs. 9(a1)–9(a3) illustrate $D_1$ dependence of the spectra of $\sigma_{sca}(\omega)$ and $F_R(\omega)$, as well as the phase of $E_{loc}$ at the HSs for asymmetric dimers while maintaining $D_2$ at 30 nm. The phase retardations of $E_{loc}$ at 90° and 180° in Fig. 9 indicate the positions of the DD and DQ coupled resonances. In the $\sigma_{sca}(\omega)$ spectra in Fig. 9(a1), both resonances shift to the lower energy region as $D_1$ increases, and the DQ coupled resonance always appears as a dip. In the $\sigma_{abs}(\omega)$ spectra in Fig. 9(a2), the absorption peak at 2.25 eV first follows the DD coupled resonance until $D_1$ reaches approximately 80–100 nm and then gradually switches to the DQ coupled resonance at 2.15 eV as $D_1$ further increases. The switch behavior is induced by the energy transfer from the DD-coupled resonance to the DQ-coupled resonance through their near-field interaction. This analysis of the $\sigma_{sca}(\omega)$ and $F_R(\omega)$ spectra, along with the phase of $E_{loc}$ at the HSs for asymmetric dimers, clarifies the experimentally observed spectral deviation of the envelopes of SERRS with SEF from the DD-coupled resonance. This deviation is induced by the excitation-light

energy transfer from the DD to the DQ coupled resonance when their spectra overlap.

To examine the mechanism of spectral deviation in the envelopes of SERRS with SEF caused by superradiant plasmon resonance in $\sigma_{sca}(\omega)$, we compared the experimental and calculated results. Fig. 10(a) shows the calculated relationship between $\hbar\omega_P$ in the $\sigma_{sca}(\omega)$ spectra and $\hbar\omega_F$ in the $F_R(\omega)$ spectra. This relationship is superimposed with the experimentally obtained relationship between $\hbar\omega_P$ and $\hbar\omega_S$ shown in Fig. 3(a). The calculated relationship reproduces both Type I and II well using the symmetric and asymmetric dimers, respectively, indicating the validity of the mechanism in which Types I and II are raised by DD and DQ coupled resonance, respectively. Fig. 10(b) presents the relationship between $\hbar\omega_P$ and $\sigma_{sca}(\omega)$ at $\hbar\omega_P$. This relationship is superimposed on the experimental relationship between $\hbar\omega_P$ at $\sigma_{sca}(\omega)$ in Fig. 3(b). Both the calculation and experimental results show the common trend of the increase in $\sigma_{sca}(\omega_P)$ causing lower energy shifts in $\hbar\omega_P$. The calculated distribution of the data for the symmetric dimers was not separated from those for the asymmetric dimers, supporting the experimental data. However, the calculated values $\sigma_{sca}(\omega_P)$ were always larger than the experimental values, indicating that the experimental NPs deviated from spherical shapes. The discussion in Fig. 10 shows that the experimentally obtained properties (1)–(4) are induced by the degree of asymmetry in the dimers.

## IV. SUMMARY

We studied the spectral relationships between SERRS with SEF and super-radiant plasmon resonance in $\sigma_{sca}(\omega)$ using a single silver NP dimer. We found two types of relationships: Type I, in which the envelopes of the SERRS with SEF are similar to the plasmon resonance maxima, and Type II, in which the envelopes shift to the higher-energy side from the maxima. Using FDTD calculations, we were able to reproduce the experimental spectra and polarization properties for both $\sigma_{sca}(\omega)$ and SERRS with SEF by changing the degree of morphological asymmetry in the dimers. Our analysis of the calculation results revealed that Types I and II are caused by DD and DQ coupled resonance, respectively. The DQ coupled resonance, which is subradiant, receives excitation light and emits SERRS and SEF light through near-field interaction with DD-coupled resonance. Our study contributes to the understanding of the EM enhancement of various plasmonic HSs composed of NP or nanowire dimers, NPs on mirrors, and NP clusters [14, 39-42]. It also shows that "Type I" and "Types II" can be uniformly described by changing the degree of contribution of subradiant resonance to SERRS with SEF and using coupling parameters between dipole and dipole or dipole and quadrupole.

**ACKNOWLEDGMENTS**

This work was supported by a JSPS KAKENHI Grant-in-Aid for Scientific Research (C) (number 21K04935).

**REFERENCES**

[1] T. Itoh, Y. S. Yamamoto, and Y. Ozaki, Plasmon-enhanced spectroscopy of absorption and spontaneous emissions explained using cavity quantum optics, Chem. Soc. Rev. **46**, 3904 (2017).

[2] S. Nie and S. Emory, Probing single molecules and single nanoparticles by surface-enhanced Raman scattering, Science **275**, 1102 (1997).

[3] K. Kneipp, Y. Wang, H. Kneipp, L. Perelman, I. Itzkan, R. R. Dasari, and M. Feld, Single molecule detection using surface-enhanced Raman scattering (SERS), Phys. Rev. Lett. **78**, 1667 (1997).

[4] H. Xu, E. J. Bjerneld, M. Käll, and L. Börjesson, Spectroscopy of single hemoglobin molecules by surface enhanced Raman scattering, Phys. Rev. Lett. **83**, 4357 (1999).

[5] A. M. Michaels, M. Nirmal, and L. E. Brus, Surface enhanced Raman spectroscopy of individual rhodamine 6G molecules on large Ag nanocrystals, J. Am. Chem. Soc. **121**,

9932 (1999).

[6] Y. S. Yamamoto, Y. Ozaki, and T. Itoh, Recent progress and frontiers in the electromagnetic mechanism of surface-enhanced Raman scattering, J. Photochem. Photobio. C **21**, 81 (2014).

[7] J. J. Baumberg, J. Aizpurua, M. H. Mikkelsen, and D. R. Smith, Extreme nanophotonics from ultrathin metallic gaps, Nat. Mater. **18**, 668 (2019).

[8] T. Itoh and Y. S. Yamamoto, Between plasmonics and surface-enhanced resonant Raman spectroscopy: toward single-molecule strong coupling at a hotspot, Nanoscale **13**, 1566 (2021).

[9] Itoh, Y. S. Yamamoto, H. Tamaru, V. Biju, S. Wakida, and Y. Ozaki, Single-molecular surface-enhanced resonance Raman scattering as a quantitative probe of local electromagnetic field: The case of strong coupling between plasmonic and excitonic resonance, Phys. Rev. B **89**, 195436 (2014); T. Itoh and Y. S. Yamamoto, Reproduction of surface-enhanced resonant Raman scattering and fluorescence spectra of a strong coupling system composed of a single silver nanoparticle dimer and a few dye molecules, J. Chem. Phys. **149**, 244701 (2018).

[10] E. C. Le Ru, P. G. Etchegoin, J. Grand, N. Félidj, J. Aubard, and G. Lévi, Mechanisms of spectral profile modification in surface-enhanced fluorescence, J. Phys.

Chem. C **111**, 44, 16076–16079 (2007).

[11] T. Itoh, Y. S. Yamamoto, H. Tamaru, V. Biju, N. Murase, and Y. Ozaki, Excitation laser energy dependence of surface-enhanced fluorescence showing plasmon-induced ultrafast electronic dynamics in dye molecules, Phys. Rev. B **87**, 235408 (2013).

[12] K. Kneipp, Y. Wang, H. Kneipp, Ir. Itzkan, R. R. Dasari, and M. S. Feld, Population pumping of excited vibrational states by spontaneous surface-enhanced Raman scattering, Phys. Rev. Lett. **76**, 2444 (1996).

[13] F. Benz, M. K. Schmidt, A. Dreismann, R. Chikkaraddy, Y. Zhang, A. Demetriadou, C. Carnegie, H. Ohadi, B. de Nijs, R. Esteban, J. Aizpurua, and J. J. Baumberg, Single-molecule optomechanics in "picocavities", Science **354**, 726 (2016).

[14] K. Yoshida, T. Itoh, H. Tamaru, V. Biju, M. Ishikawa, and Y. Ozaki, Quantitative evaluation of electromagnetic enhancement in surface-enhanced resonance Raman scattering from plasmonic properties and morphologies of individual Ag nanostructures, Phys. Rev. B **81**, 115406 (2010).

[15] T. Itoh, M. Iga, H. Tamaru, K. Yoshida, V. Biju, and M. Ishikawa, Quantitative evaluation of blinking in surface enhanced resonance Raman scattering and fluorescence by electromagnetic mechanism, J. Chem. Phys. **136**, 024703 (2012).

[16] M. D. Doherty, A. Murphy, R. J. Pollard, and P. Dawson, Surface-enhanced Raman

scattering from metallic nanostructures: bridging the gap between the near-field and far-field responses, Phys. Rev. X **3**, 011001 (2013).

[17] S. L. Kleinman, B. Sharma, M. G. Blaber, A.-I. Henry, N. Valley, R. G. Freeman, M. J. Natan, G. C. Schatz, and R. P. Van Duyne, Structure enhancement factor relationships in single gold nanoantennas by surface-enhanced Raman excitation spectroscopy, J. Am. Chem. Soc. **135**, 301−308 (2013).

[18] Y. Tanaka, A. Sanada, and K. Sasaki, Nanoscale interference patterns of gap-mode multipolar plasmonic fields, Sci. Rep. **2**, 764 (2012); M. Liu, T.-W. Lee, S. K. Gray, P. Guyot-Sionnest, M. Pelton, Excitation of dark plasmons in metal nanoparticles by a localized emitter, Phys. Rev. Lett. 102, 107401 (2009).

[19] J. Ye, F. Wen, H. Sobhani, J. B. Lassiter, P. Van Dorpe, P. Nordlander, N. J. Halas, Plasmonic nanoclusters: near field properties of the Fano resonance interrogated with SERS, Nano Lett. **12**, 1660 (2012).

[20] S. Simoncelli, Y. Li, E. Cortes, and S. A. Maier, Imaging plasmon hybridization of Fano resonances via hot-electron-mediated absorption mapping, Nano Lett. **18**, 3400−3406 (2018).

[21] S. Chen, Y. Zhang, T.-M. Shih, W. Yang, S. Hu, X. Hu, J. Li, B. Ren, B. Mao, Z. Yang, and Z. Tian, Plasmon-induced magnetic resonance enhanced Raman spectroscopy, Nano

Lett., **18**, 2209−2216 (2018).

[22] G.-C. Li, Y.-L. Zhang, J. Jiang, Y. Luo, and D. Y. Lei, Metal-substrate-mediated plasmon hybridization in a nanoparticle dimer for photoluminescence line-width shrinking and intensity enhancement, ACS Nano, 11, 3067−3080 (2017).

[23] B. Gallinet and O. J. F. Martin, Influence of electromagnetic interactions on the line shape of plasmonic Fano resonances, ACS Nano **5**, 8999–9008 (2011).

[24] R. Adato, A. Artar, S. Erramilli, and H. Altug, Engineered absorption enhancement and induced transparency in coupled molecular and plasmonic resonator systems, Nano Lett. **13**, 2584−2591 (2013).

[25] J. Langer, et al., Present and future of surface-enhanced Raman scattering, ACS Nano **14**, 28-117 (2020).

[26] P. Lee and D. Meisel, Adsorption and surface-enhanced Raman of dyes on silver and gold sols, J. Phys. Chem. **86**, 3391 (1982).

[27] E. C. Le Ru, M. Meyer, and P. G. Etchegoin, Proof of single-molecule sensitivity in surface enhanced Raman scattering (SERS) by means of a two-analyte technique, J. Phys. Chem. B **110**, 1944 (2006).

[28] A. B. Zrimsek, A. I. Henry, and R. P. Van Duyne, Single molecule surface-enhanced Raman spectroscopy without nanogaps, J. Phys. Chem. Lett. **4**, 3206 (2013).

[29] T. Itoh, Y. S. Yamamoto, and T. Okamoto, Absorption cross-section spectroscopy of a single strong-coupling system between plasmon and molecular exciton resonance using a single silver nanoparticle dimer generating surface-enhanced resonant Raman scattering, Phys. Rev. B, **99**, 235409 (2019).

[30] J. A. Fan, C. Wu, K. Bao, J. Bao, R. Bardhan, N. J. Halas, V. N. Manoharan, P. Nordlander, G. Shvets, and F. Capasso, Self-assembled plasmonic nanoparticle clusters, Science **328**, 1135 (2010).

[31] Z. J. Yang, Z. S. Zhang, L. H. Zhang, Q. Q. Li, Z. H. Hao, and Q. Q. Wang, Fano resonances in dipole-quadrupole plasmon coupling nanorod dimers, Opt. Lett. **36**,1542, (2011).

[32] C. F. Bohren and D. R. Huffman, *Absorption and Scattering of Light by Small Particles* (Wiley, New York, 1983).

[33] P. B. Johnson and R. W. Christy, Optical constants of the noble metals, Phys. Rev. B **6**, 4370–4379 (1972).

[34] E. Elliott, K. Bedingfield, J. Huang, S. Hu, B. de Nijs, A. Demetriadou, and J. J Baumberg, Fingerprinting the hidden facets of plasmonic nanocavities, ACS Photonics 9, 2643−2651 (2022).

[35] P. Johansson, H. Xu, and M. Käll, Surface-enhanced Raman scattering and

fluorescence near metal nanoparticles, Phys. Rev. B 72, 035427 (2005).

[37] L. V. Brown, H. Sobhani, J. B. Lassiter, P. Nordlander, and N. J. Halas, Heterodimers: plasmonic properties of mismatched nanoparticle pairs, ACS Nano **4**, 819–832 (2010).

[38] C. Sönnichsen, T. Franzl, T. Wilk, G. von Plessen, J. Feldmann, O. Wilson, and P. Mulvaney, Drastic reduction of plasmon damping in gold nanorods, Phys. Rev. Lett. **88**, 077402 (2002).

[39] T. Itoh, T. Y. S. Yamamoto, Y. Kitahama, and J. Balachandran, One-dimensional plasmonic hotspots located between silver nanowire dimers evaluated by surface-enhanced resonance Raman scattering, Phys. Rev. B **95**, 115441 (2017).

[40] T. Itoh, T. Y. S. Yamamoto, and J. Balachandran, Propagation mechanism of surface plasmons coupled with surface-enhanced resonant Raman scattering light through a one-dimensional hotspot along a silver nanowire dimer junction, Phys. Rev. B **103**, 245425 (2021).

[41] C. Ciraci, R. T. Hill, J. J. Mock, Y. Urzhumov, A. I. Fernandez-Dominguez, S. A. Maier, J. B. Pendry, A. Chilkoti, and D. R. Smith, Probing the ultimate limits of plasmonic enhancement, Science **337**, 1072–1074 (2012).

[42] S. Y. Ding, E. M. You, Z. Q. Tian, and M. Moskovits, Electromagnetic theories of surface-enhanced Raman spectroscopy, Chem. Soc. Rev. **46**, 4042–4076 (2017).

**Figure captions**

 Dark-field illumination of a silver NP dimer. The dimer placed on the cover glass is excited from above and forward elastic scattered light is detected.  Dark-field image of NPs and NP aggregates on the cover glass. The single light spot is selected by a pinhole to measure the elastic scattering spectra of a single dimer. The scale bar is 5 μm.  SERRS and SEF excitation of a NP dimer. The dimer is excited from above and forward SERRS and SEF light is detected.  SERRS and SEF image of NP aggregates on the cover glass. The scale bar is 5 μm.

 $\sigma_{\text{sca}}(\omega)$ spectra (blue lines) and spectra of SERRS and SEF (red lines) of single dimers exhibiting similar spectral shapes (Type I).  $\sigma_{\text{sca}}(\omega)$ spectra and spectra of SERRS and SEF for single dimers exhibiting spectral deviation with respect to each other (intermediate type).  $\sigma_{\text{sca}}(\omega)$ spectra and spectra of SERRS and SEF with maxima that are close to spectral dips in the elastic scattering (Type II).

 Relationship between the $\hbar\omega_{\text{p}}$ and $\hbar\omega_{\text{s}}$ of single dimers exhibiting Type I

(black circles) and Type II (red circles). (b) Relationship between the $\hbar\omega_p$ and $\sigma_{sca}(\omega_p)$ of single dimers exhibiting Type I (black circles) and Type II (red circles).

FIG. 4(a1)–4(a4) SERRS with SEF (red lines) and $\sigma_{sca}(\omega)$ spectra (blue lines) of dimers, respectively. Insets are SEM images of the dimers. (b1)–(b4) SERRS with SEF (red lines) and $\sigma_{sca}(\omega)$ spectra (blue lines) of dimers, respectively. Insets are SEM images of the dimers. Scale bars are 100 nm.

FIG. 5 (a) FDTD calculation setup for electric fields around a single NP dimer by wide-field excitation from the upper side with respect to the coordinate system. Polarization directions of the excitation light are indicated by gray arrows. (b) Dimer composed of two NP with diameters of $D_1$ and $D_2$, respectively. The gap was set to 1 nm. The position of the HS is indicated by the red dot (c) $\sigma_{sca}(\omega)$ spectrum (blue curve) of a dimer with $D_1$ and $D_2$ of 35 nm with the incident light polarized with respect to the x-axis. (d) $F_R(\omega)$ spectrum (red curve) of a dimer with $D_1$ and $D_2$ of 35 nm with incident light polarized with respect to the x-axis with a phase of $E_{loc}$ (black curve) at the center gap, indicated by the red dot in (b).

FIG. 6 (a1)–(a10) $\sigma_{sca}(\omega)$ spectra of single NPs with diameters of 30, 35, 40, 50, 60, 80, 100, 120, and 140 nm, respectively. D, Q, and H indicate dipole, quadrupole, and hexapole resonance, respectively. (b1) and (b2) Dipole and quadrupole on single NPs raised by light coming along the y axis. (c1) and (c2) Dipole and quadrupole coupled resonance for symmetric and asymmetric dimers, respectively.

FIG. 7 (a1)–(a6) $\sigma_{sca}(\omega)$ spectra (blue lines) and $F_R(\omega)$ spectra (red lines) of single symmetric dimers with diameters of 30, 40, 60, 80, 100, and 120 nm, respectively. (b1)–(b6) $\sigma_{sca}(\omega)$ spectra (blue lines) and $F_R(\omega)$ spectra (red lines) of single asymmetric dimers with a $D_1$ of 30 nm and $D_2$ of 30, 50, 80, 100, 120, and 140 nm, respectively.

FIG. 8 (a1)–(a4) $D_1$ dependence of the $\sigma_{sca}(\omega)$ spectra with the phases of $E_{loc}$ with a $D_1$ of 30 nm and $D_2$ of 30, 50, 80, and 120 nm, respectively. The initial phase of $E_I$ and $D_2$ were set 180° and 30 nm, respectively. (b1)–(b4) $D_1$ dependence of the $F_R(\omega)$ spectra and the phases of $E_{loc}$ with a $D_1$ of 30 nm and $D_2$ of 30, 50, 80, and 120 nm, respectively. The initial phase of $E_I$ and $D_2$ were set 180° and 30 nm, respectively.

FIG. 9 (a1)–(a3) $D_1$ dependences of spectra of $\sigma_{sca}(\omega)$ (a1) and $F_R(\omega)$ (a2) as well as the phase of $E_{loc}$ (a3) at HSs for asymmetric dimers while fixing $D_2$ at 30 nm, expressed as contour maps. The black circles and triangles indicate the phase retardations of $E_{loc}$ for 90° and 180°, respectively, at HSs in Fig. 6(b). The higher energy regions in (a3) are sealed to avoid complex phase inversion.

FIG. 10 (a) Calculated relationship between $\hbar\omega_p$ and $\hbar\omega_F$ of single symmetric dimers (○) and asymmetric dimers with $D_2$ values of 35 (△), 40 (▽), 60 (◇) nm. The diameters of the NP symmetric dimers are 30–140 nm. $D_1$ values for the asymmetric dimers are 35–140 (△), 40–140 (▽), and 60–140 (◇) nm. The experimental relationship between $\hbar\omega_p$ and $\hbar\omega_S$ of single dimers (○) taken from Fig. 3(a). (b) Calculated relationship between $\hbar\omega_p$ and $\sigma_{sca}(\omega_p)$ of single symmetric dimers (○) and asymmetric dimers with $D_2$ values of 35 (△), 40 (▽), and 60 (◇) nm. The diameters of the NP symmetric dimers are 30–140 nm. $D_1$ values for the asymmetric dimers are 35–140 (△), 40–140 (▽), 60–140 (◇) nm, respectively. The experimental relationship between $\hbar\omega_p$ and $\sigma_{sca}(\omega_p)$ of single dimers (○) from Fig. 3(b).

## Figure 1

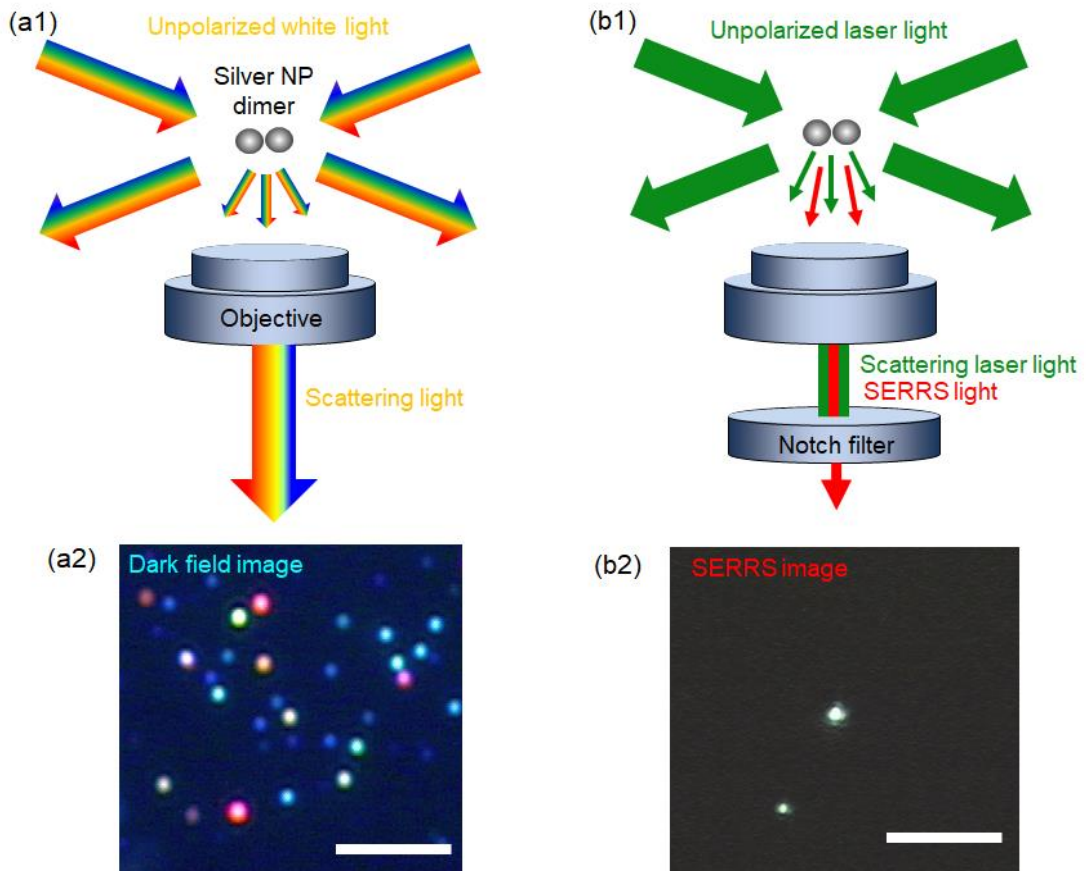

Figure 2

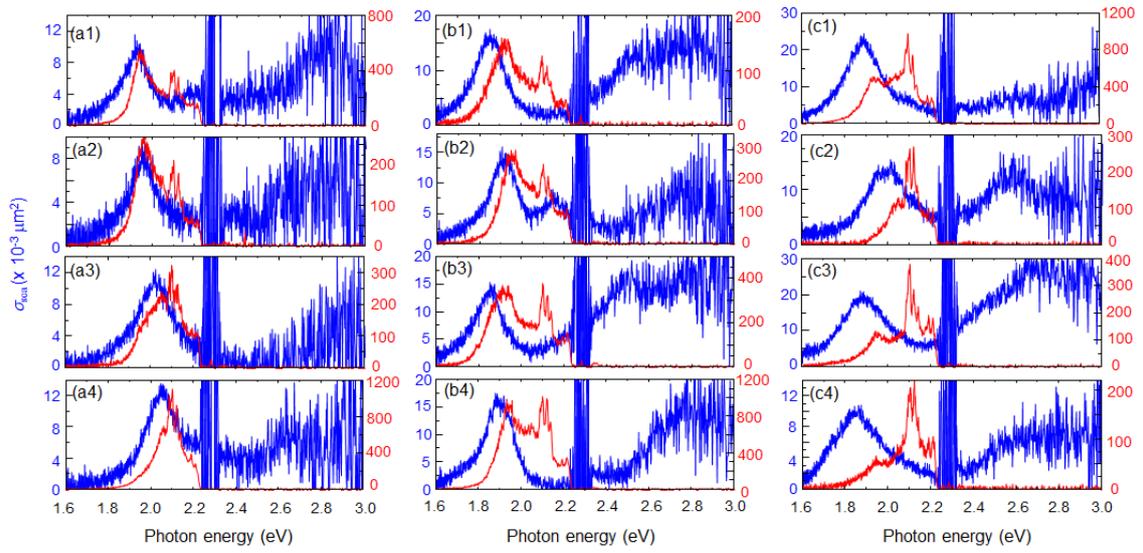

# Figure 3

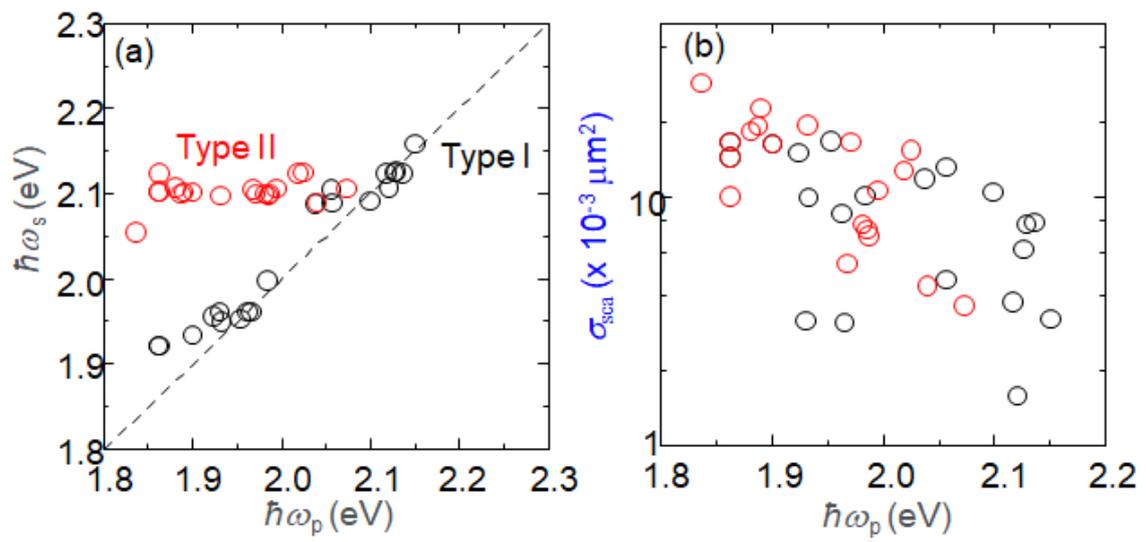

## Figure 4

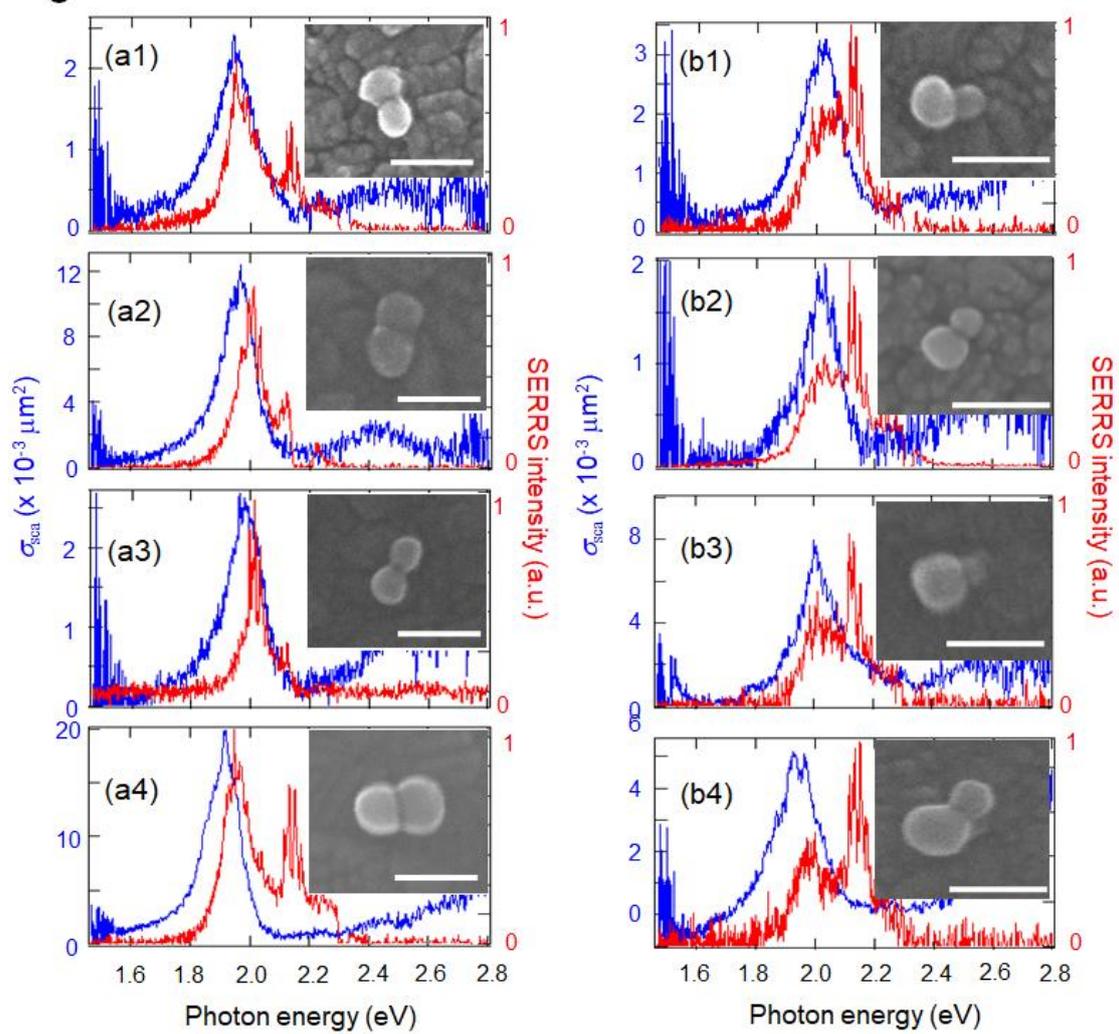

# Figure 5

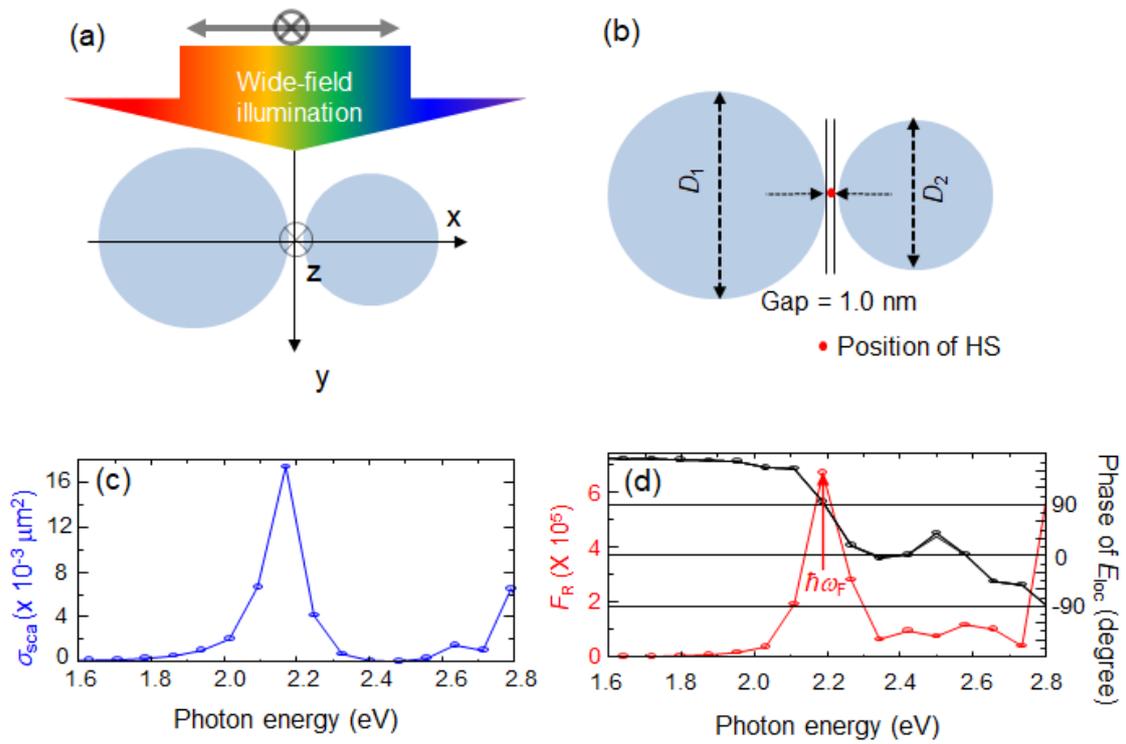

# Figure 6

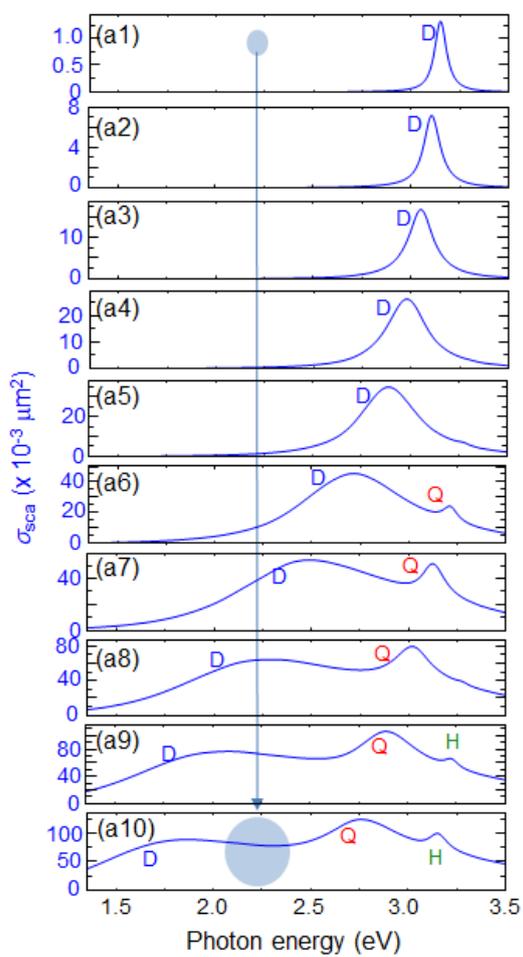

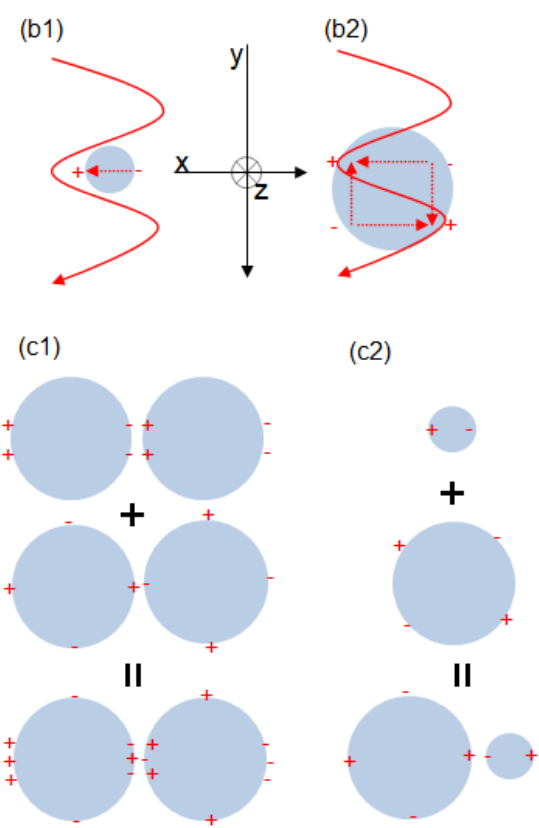

## Figure 7

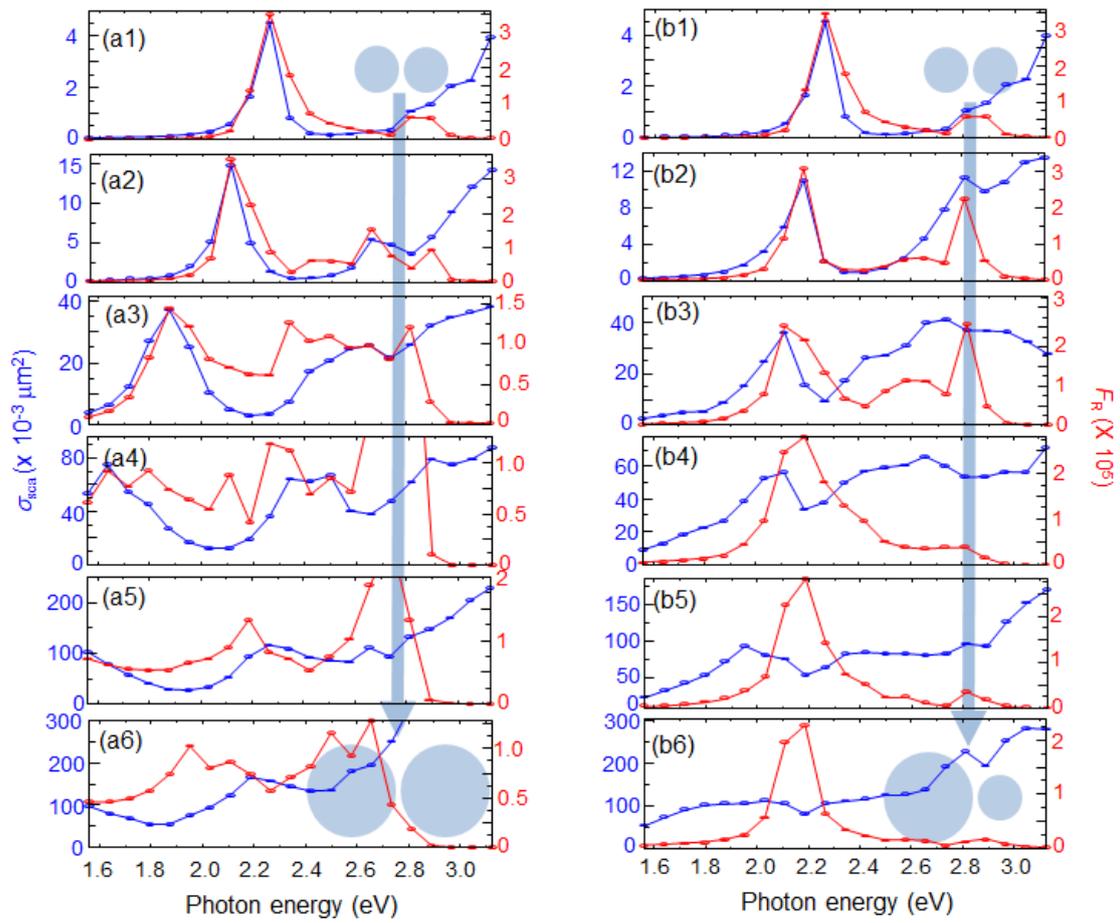

Figure 8

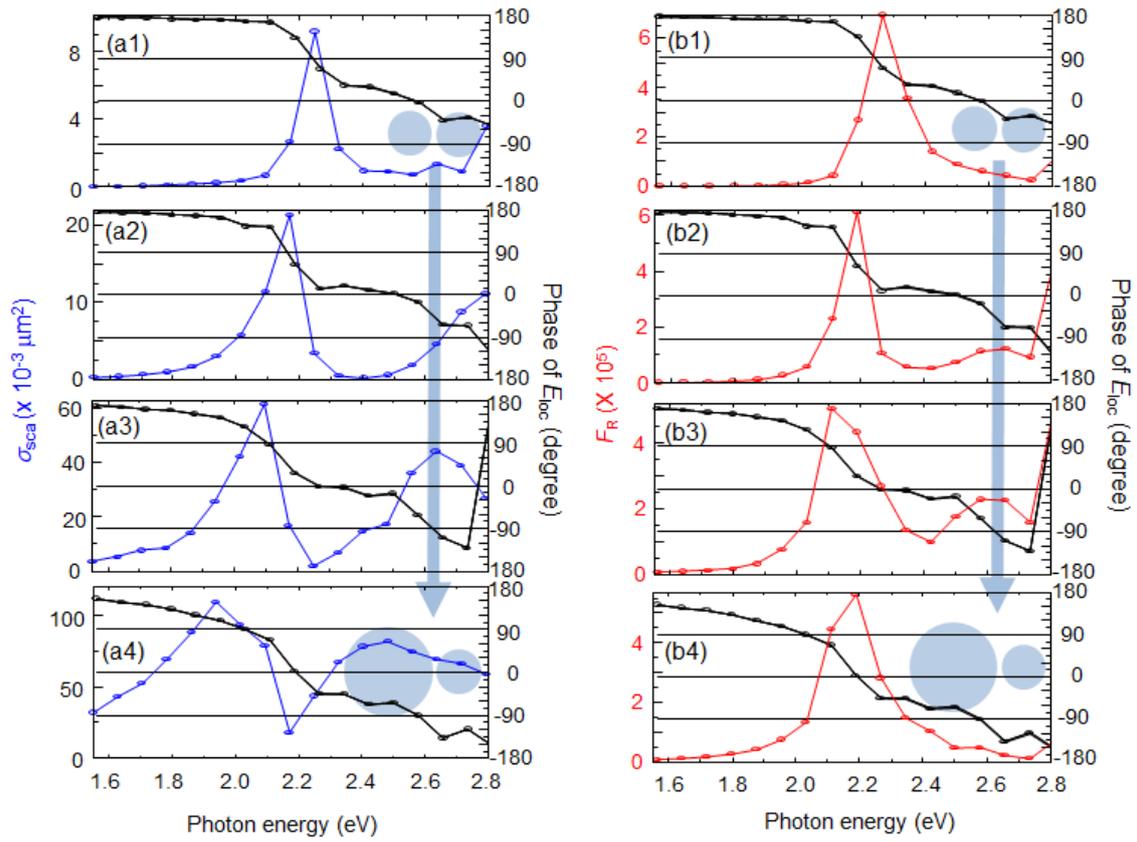

# Figure 9

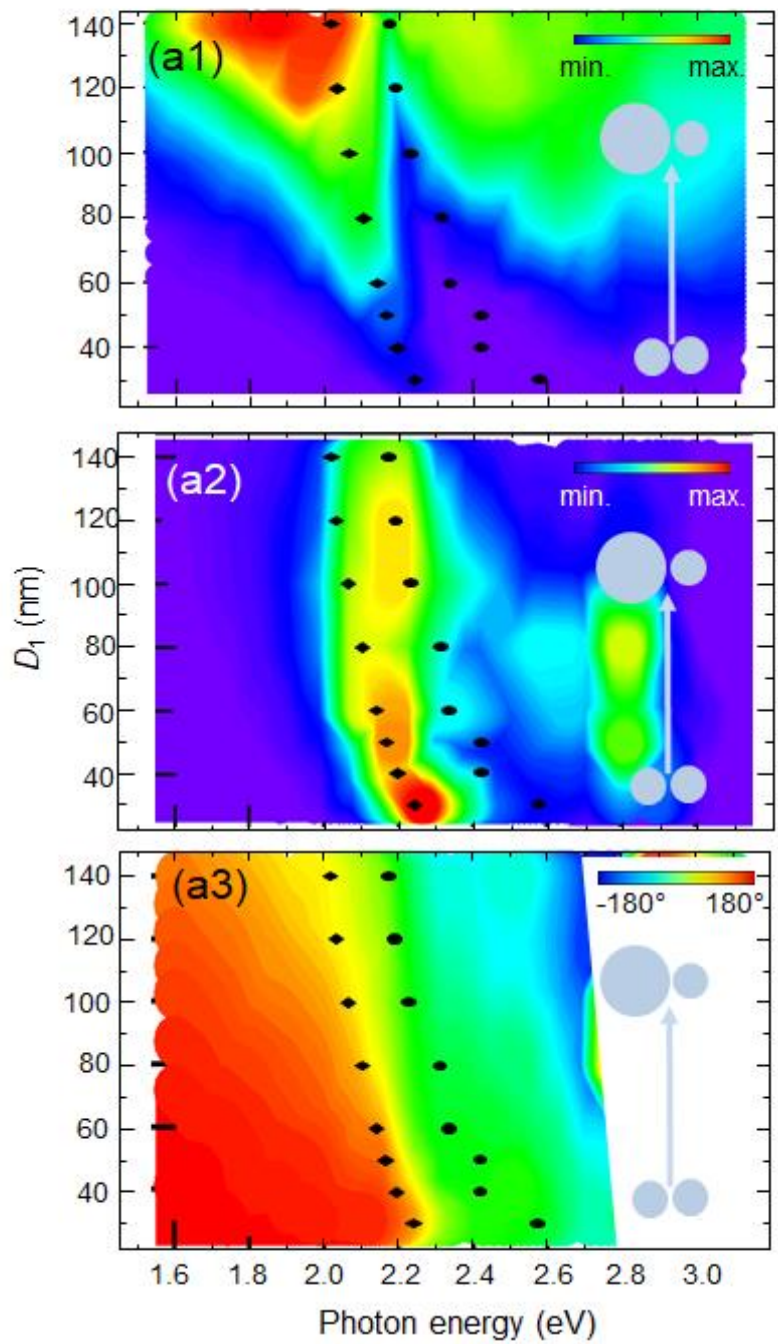

# Figure 10

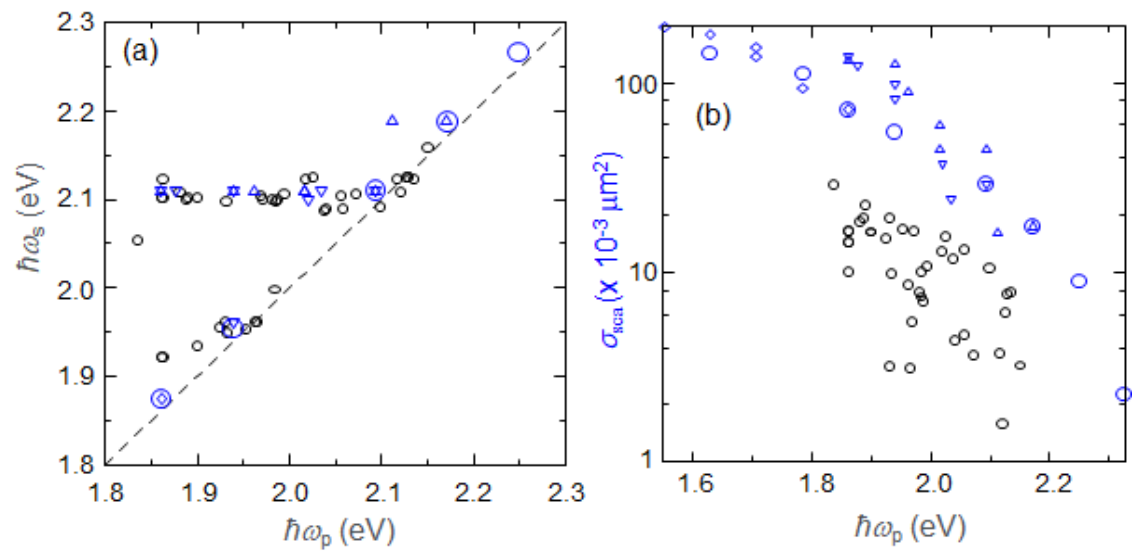